\documentclass[twocolumn,pra,aps,showpacs]{revtex4}
\usepackage{graphicx}
\usepackage{amsmath}
\usepackage{amssymb}

\begin{document}

\title{Deriving Landauer's erasure principle from statistical mechanics}

\author{Kurt Jacobs}

\affiliation{Quantum Science and Technologies Group, Hearn Institute for Theoretical Physics, Department of Physics and Astronomy, 202 Nicholson Hall, Tower Drive, Baton Rouge, LA 70803, USA }
\affiliation{Centre for Quantum Computer Technology, Centre for 
Quantum Dynamics, School of Science, Griffith University, Nathan 4111, 
Brisbane, Australia}

\begin{abstract}
We present a concise derivation of Landauer's erasure principle from the postulates of statistical mechanics, along with a small number of additional but uncontroversial axioms. 
\end{abstract}


\maketitle

Landauer's erasure principle~\cite{L61} states that whenever a single bit of information is erased, the entropy in the environment to which the information storing system is connected must increase by at least $k\ln 2$, where $k$ is Boltzmann's constant. (This principle is also often stated by saying that the energy of the environment must increase by $kT\ln 2$, which is equivalent to the former statement for a canonical reservoir at temperature $T$.) Landauer's erasure principle is an important tool in understanding thermodynamics, particularly Maxwell's demon~\cite{B82,LR90,B87,B02,Bub}, and in understanding the relationship between thermodynamics and classical and quantum information theory~\cite{V,PV}. A comprehensive review of the literature on Landauer's erasure and Maxwell's demon may be found in~\cite{LR03}.

Here we present a concise, but rigorous derivation of Landauer's erasure principle from the postulates of statistical mechanics, along with small set of other fundamental postulates, all of which are uncontroversial. Certainly it has been discussed previously how Landauder's principle follows from the second law of thermodynamics~\cite{L61,B82,L87,Z84,LR90}, and the present treatment is offered primarily in the hope that it is more transparent than its predecessors. Previous mathematical derivations of Landauer's principle have usually focussed on example systems~\cite{L61,S95,P00}. While the treatment in~\cite{P00} does apply to a wide class of systems, it is simpler to proceed directly from the basic postulates of statistical mechanics, and the resulting derivation is no less rigorous. It is hoped that this treatment will at the least provide a useful pedagogical aid, and make it crystal clear that Landauer's principle is as general as those of statistical mechanics. If one wishes to argue against the applicability of Landauer's principle, one must argue against the applicability of at least one of the postulates of statistical mechanics. 

Before we begin we must define what is meant by ``erasing the information" contained in a system ${\mathcal S}$. What is meant is a process by which all the possible states of ${\mathcal S}$ are mapped to a single state. Thus, after such a process,  one no longer knows what state the system was in before, and the information has therefore been erased.

It is important to understand that to perform such an erasure operation we must ultimately use a process which is invertible --- that is, we must use a process that is logically reversible. This is because the fundamental laws of physics are reversible, and as such reversible operations are the only operations that are available to us. (Quantum mechanically this means that the erasure operation must be performed with a unitary transformation.) However, as stated this would appear to lead a contradiction, since the definition of erasure we have just given is that two or more states are mapped to a single state, a operation which is clearly not invertible. What we {\em actually} mean by erasure then is the following: we apply a reversible operation which takes the many initial states of the system and maps them to different  microstates of an environment so that these states are no longer accessible to us. Having achieved this, the system is left in a single state, and since we have no access to the microstates of the environment, we no longer have any information about the initial state of the system.  As a result we say that the information has been erased. The use of an environment is thus essential for the process of erasure. 

To begin with we present a simple argument, which while persuasive is nevertheless insufficient to derive Landauer's principle.  (To make this argument concise I assume a knowledge of quantum mechanics. Readers unfamiliar with quantum mechanics may skip this paragraph.). Consider the physical system containing the information we wish to erase, ${\mathcal S}$, to be described by the density matrix $\rho$. The amount of information which ${\mathcal S}$ contains is given by the von Neumann entropy $S_{vN}(\rho)$. If we erase the information in $S$ by interacting the system with an environment, ${\mathcal E}$, then we must employ a unitary operation on the joint state-space of ${\mathcal S}$ and ${\mathcal E}$. The system and environment are initially separate, and a result the entropy of the joint system is simply $S_{vN}(\rho) + S_{vN}(\sigma)$  where $\sigma$ is the state of the environment. Since the unitary interaction  cannot change the entropy of the total system, and the final entropy of ${\mathcal S}$ is zero (because it is in the single pure state ($|0\rangle$), the von Neumann entropy of the environment must have increased by $S_{vN}(\rho)$, which is the  amount of the information which has been erased. Since the thermodynamic entropy is given by $S = kS_{vN}$, where $k$ is Boltzmann's constant, the entropy of the environment must increase by $k S_{vN}(\rho)$. If  the system initially contains one bit of information, then $S_{vN}(\rho) = \ln 2$,  so that the entropy of the environment increases by $k\ln 2$.

The above argument requires that we equate the von Neumann entropy with the thermodynamic entropy, and will thus leave many people uneasy. However, the argument is incomplete for another reason. It implies that the increase in entropy of the environment depends on the initial state of the system. In fact, the entropy in crease has nothing to do with the initial state of the system~$\rho$~\cite{L61,B82,LR03}. It is actually a result of the fact that the transformation which achieves the erasure is non-invertible. In particular, ${\mathcal S}$ can be in a pure state, and the increase in the entropy of the environment is still $k\ln M$,  where $M$ is the dimension of the system. This is clear because if the system starts in a mixed state, say an equal probability of being in the two states $|0\rangle$ or $|1\rangle$, this simply means that the system is randomly prepared in state $|0\rangle$  or $|1\rangle$. Thus on any given run the system has a definite state, and one would expect Landauer's principle to hold true for every run. 

To derive Landauer's principle, and in doing so see that the entropy gain of the reservoir is independent of the initial state of the system (i.e., independent of what information the system holds), we will need to use one of the postulates of statistical mechanics. Statistical mechanics arose out of the observation that once the macroscopic  properties of a system are specified (that is, the macroscopically observable state-variables such as energy, pressure etc.), there are very many possible physical states that that system could be in. Two of the postulates of statistical mechanics are 1. That each of the possible physical states, being those which are consistent with the macroscopic properties of the system) are {\em equally likely}, and 2. That the thermodynamic entropy is given by $k\ln N$, where $N$ is the total number of possible physical states~\cite{SM}. It is the second of these which will lead to Landauer's principle. We quote postulate 1 merely because it shows the connection between information theoretic entropy and statistical mechanical entropy: if all the microstates are equally likely, then the information theoretic entropy of the system is $\ln N$. However, this connection is not required in what follows. 

We will stick to classical systems in the following derivation because the use
of quantum systems merely complicates the language without adding anything 
important. Without loss of generality we will take our system to have two 
states, and label these states with the binary digits 0 and 1. The environment,
or reservoir, has a great many microstates. We will label each of these states
with a unique number, and write this number in trinary. Thus each state will
be represented  as a long string of trinary digits, each digit having the value
0, 1 or 2.  (Trinary is not essential, but will be convenient.) 

At the start of the process the system may be in either of its two possible states. We will give the environment an initial entropy of  $k\ln N$. This means that, while the environment does exist in some definite microstate, there are $N-1$ other microstates which give the same value for its macroscopic state-variables~\cite{note}. Thus, at the start of the procedure the environment is allowed to be in any one of a set of $N$ possible states. Considering our trinary representation for each of these states, some of the trinary digits (trits) will be the same for all the $N$ states, but the value of many of them will depend upon which of the $N$ states the system is in. 

As explained above, to perform the erasure of the system, one effects a reversible transformation coupling the system to the environment. Further, this reversible transformation is fixed - that is, it is independent of the initial state of the system, and independent  of which of the $N$ states the environment is in. The reason for this will be explained later. Recall that the reversibility of the operation implies that it must be possible to determine the original state of the system and environment from the  final state -- put another way, the transformation must be invertible. 

Now we apply the erasure condition, which states that the system must be in the state 0 after the process. Combining this with the fact that the operation must be invertible, means that the environment must record which state the system was initially in, since the system does not. This can be done by mapping one of the trits of the environment to state 0 if the system  was in state 0, and 1 if it was in state 1. However, now the initial state  of the trit must also be recorded somewhere for the process to be invertible. If we record the state of this trit in another trit, we are again faced with the same problem. We cannot store the state of the trit in the  {\em system}, because the final state of the system must be 0. The consequence of all this is that the initial state of the trit in which we will store the initial state of the system must be fixed at the start so that we do not need to store its value anywhere. We will take the initial value of this trit to be 2. 

To sum up the above discussion, we see that the process of erasure takes  the system to a fixed state, and one of the trits which {\em was initially} in a fixed state, to one of two states. (One can also imagine that two trits are used instead, or that an initially fixed trit is used to store the initial  state of a different trit which is altered, but the result is the same.)  At the end of the procedure, the environment is now in one of $2N$ possible states,  precisely because a trit that was initially fixed, is now in one of two states. 

Finally, to obtain Landauer's erasure principle, we apply the erasure condition again: If the information in the system is to be erased, then the macroscopic observables must not depend on the initial state of the system (otherwise we would still have access to the information we are trying to erase). Thus, the macroscopic state-variables must be the same for all of these $2N$ states. Each of these states is thus a member of a set of at least $2N$ states for which the macroscopic state-variables are the same. As a result, the final entropy of the environment is at least $k\ln(2N) = k\ln N + k\ln2$, and has therefore increased by at least $k\ln 2$.

Since the final states are members of a set of $2N$ states for which the macroscopic state-variables are the same, whereas the initial states are members of a set of just $N$ such states, the final states must all be different from the initial states. It is now clear why we chose the initial state of the trit which is altered as 2, and the final states as 0 and 1. If we had  chosen the final states to be 1 and 2, respectively, so that the state of  the environment was not altered by the erasure when the system started in the 0 state, then the following would be true: When the system started  in the zero state, the environment would remain in the set of $N$ states  for which the macroscopic state-variables have their initial values. However, when the  system started in the 1 state, the state of the environment would be mapped  {\em outside} this set of states. As a result, the final values of the macroscopic state variables would record the initial state of the system, and the information would not have been erased.

It is now clear that regardless of the initial state of the system, the entropy 
of the environment must increase, and what is more the minimum increase is the
{\em same} for all initial states.

In the above derivation we employed the fact that the reversible transformation 
used to erase the information is fixed, by which we mean that it cannot 
depend upon the initial state of the system or the environment. We now explain 
why this is the case. Most fundamentally this follows from the assumption 
that the universe is evolving according to the laws of physics, and thus 
obeying an evolution due to a single reversible transformation. Thus there 
is only one transformation which does everything in the universe, including, 
necessarily, erasing the information in our system. 

However, from everyday experience we know that we could, if we chose, perform one of two unitary transformations depending on the initial state of the system (we use unitary transformations here because we assume the that universe in quantum mechanical). It is therefore worth examining how this is done. It must be accomplished with a single unitary, and as result we require is a single unitary in a larger state space, which contains the two unitaries as sub-matrices. Each of the sub-matrices is an operator which acts on our 
system, and so is two dimensional. The full unitary therefore is four dimensional as it contains two of these operators. It therefore acts in a four dimensional space, and we choose this to be the tensor product of two two-state systems. It is the state of the second ``auxiliary'' system which determines which unitary will act on our system. We wish this to depend on the state of our system, so we perform a unitary between our system and the auxiliary system which correlates the two - that is, sets the auxiliary to $|1\rangle$ of the 
system is in state $|1\rangle$, etc. Then we evolve the single unitary operation, which enacts either of the two unitaries on our system, and is thus able to prepare the system in state $|0\rangle$ regardless of its initial state. So what's the catch? The catch is that the auxiliary system now records the initial state of our system, and the information has therefore not been erased. To put this more simply another way, if we use two unitaries 
to ``erase'' the state of the system, as we know we can, then we know which 
unitary we used and therefore the information has not, in fact, been erased. 

To sum up the above analysis, Landauer's erasure principle follows from four basic 
postulates: the definition of erasure, the statistical mechanical definition of entropy, the necessity of using a reversible transformation to perform the erasure, and finally the fact that the unitary operation which is used must be independent of the state of the system. 

{\em Acknowledgments}: This work was supported by the Australian Research Council, the State of Queensland, The Hearne Institute, The National Security Agency, The Army Research Office and The Disruptive Technologies Office.


\begin{thebibliography}{99}
\bibitem{note}
   Here we talk as if the macroscopic state variables are precisely defined. Of course these variables are only known to some specified precision. To be precise, the $N$ possible microstates of the system are those that are consistent with the macroscopic variables to the precision with which they are known. We avoid specific reference to this in the text so as to keep the language uncluttered. Thus, when we say in the text that the macroscopic variables are the same for two different microstates, we mean that they are the same {\em within the error to which they are known}.
\end{thebibliography}
\end{document}